\documentclass[journal=jacsat,manuscript=article]{achemso} 

\usepackage{chemformula} 
\usepackage[T1]{fontenc} 
\usepackage{booktabs}
\usepackage{multirow}
\usepackage{siunitx}
\usepackage{caption}
\usepackage{threeparttable}
\usepackage{array}
\usepackage{amsmath}
\usepackage{amsfonts}
\usepackage{hyperref}
\usepackage{amsmath}
\usepackage{graphicx}
\usepackage{mathtools}
\usepackage{multicol}
\usepackage{amssymb}
\usepackage{threeparttable}
\usepackage{adjustbox}
\usepackage{array}
\usepackage{siunitx}
\usepackage{hyperref}
\usepackage[compact]{titlesec}
\usepackage{parskip}
\usepackage{tocloft}
\usepackage{booktabs}

\author{Xinyu Xu}
\affiliation[Anhui]
{School of Physics, Anhui University, Hefei 230601, China}

\author{Arif Ullah}
\email{arif@ahu.edu.cn}
\affiliation[Anhui]
{School of Physics, Anhui University, Hefei 230601, China}

\author{Ming Yang}
\email{mingyang@ahu.edu.cn}
\affiliation[Anhui]{School of Physics, Anhui University, Hefei 230601, China}

\title[An \textsf{achemso} demo]
  {A Physics-Informed Chemical Rule for Topological Materials Discovery}
\abbreviations{IR,NMR,UV}

\begin{document}

\begin{abstract}
Topological phases of matter—comprising both insulators and semimetals—offer great potential for quantum applications, but identifying new candidates remains challenging due to expensive first-principles simulations and labor-intensive experimental workflows. Here we introduce a physics-informed chemical rule that integrates compositional, orbital and crystallographic descriptors within an interpretable linear framework. By explicitly encoding electron filling, space-group symmetry and orbital-resolved chemical environments, our method overcomes a fundamental limitation of composition-only heuristics—their inability to distinguish polymorphs with identical stoichiometry but different crystal structures. Using only elemental characteristics, our approach reduces a material's topological propensity to a single, physically interpretable score, enabling rapid and high-throughput assessment. The model achieves superior predictive performance while maintaining physical transparency, and identifies candidate topological materials where conventional symmetry indicators fail. Consequently, our framework enables rapid and interpretable exploration of complex materials spaces, establishing a scalable paradigm for the intelligent discovery of next-generation topological and quantum materials.

\end{abstract}

\section{Introduction}

Topological materials, encompassing topological insulators (TIs)\cite{hasan2010topological, qi2011topological} and topological semimetals (TSMs),\cite{liu2014discovery, xu2015discovery} represent a broad class of quantum matter distinguished by nontrivial electronic band topology characterized by quantized invariants.\cite{haldane1988model,kane2005Z2,armitage18weyl} TIs feature an insulating bulk alongside symmetry-protected conducting surface states,\cite{kane2005quantum,bernevig2006quantum,konig2007quantum,fu2007topological,hsieh2008atopological,fu2011topological,hsieh2012topological,benalcazar2017quantized} while TSMs host gapless excitations such as Weyl or Dirac nodes accompanied by nontrivial Berry curvature distributions.\cite{lv2015experimental,burkov2011topological,bradlyn2016beyond} The robust boundary modes emerging from these phases give rise to exotic phenomena—including the quantum spin Hall effect,\cite{kane2005quantum} anomalous transport signatures,\cite{qi2011topological} and magnetoelectric responses\cite{hu2019transport}—positioning topological materials as compelling platforms for next-generation quantum, electronic, and spintronic technologies. A central challenge in the field has therefore been the reliable identification and classification of such materials.

Early efforts to identify topological materials relied heavily on first-principles electronic structure calculations integrated with topological band theory,\cite{xiao2021first, bansil2016band} a computationally intensive yet rigorous approach for establishing topological order. A major breakthrough came with the development of symmetry indicators\cite{slager2013space, kruthoff2017topological, po2017symmetry} and topological quantum chemistry (TQC),\cite{bradlyn2017topological} which enabled efficient diagnosis of topological phases directly from symmetry representations of electronic states. By linking symmetry representations at high-symmetry points in the Brillouin zone to global band topology, these frameworks have facilitated high-throughput computational screening, resulting in extensive databases of candidate topological materials and accelerating both theoretical predictions and experimental validation.\cite{vergniory2019complete, zhang2019catalogue, tang2019comprehensive}

Despite their success, symmetry-based methodologies are not universally applicable. Their reliance on well-defined crystalline symmetries restricts their ability to capture topological phases arising from symmetry-breaking effects, accidental band inversions, or strong spin–orbit coupling beyond perturbative regimes.\cite{po2017symmetry} Certain phases—such as Chern insulators and time-reversal-invariant $\mathbb{Z}_2$ insulators lacking point group symmetries—remain inaccessible to symmetry indicator diagnostics, requiring instead explicit evaluation of wave-function-based topological invariants, a computationally demanding process.\cite{po2017symmetry} A notable example is the Weyl semimetal TaAs,\cite{xu2015discovery,lv2015experimental} whose topological character cannot be deduced from symmetry indicators alone but requires explicit evaluation of Berry curvature and Chern numbers.\cite{weng2015weyl} Materials with low crystal symmetry or complex magnetic ordering present additional challenges for symmetry-based classification.\cite{xu2020high} These constraints underscore a fundamental gap: the absence of a general and physically transparent framework capable of identifying topological character across materials with diverse symmetries, compositions, and electronic structures.

In recent years, machine learning (ML) has emerged as a scalable alternative to symmetry-based approaches for classifying and predicting topological materials properties.\cite{zhang2018machine, scheurer2020unsupervised, cao2020artificial, choudhary2021high, schleder2021machine, tyner2024machine, hong2025discovery, haosheng2025predicting, he2025machine} By learning relationships between material descriptors and topological properties, ML models can bypass expensive wave-function calculations and enable rapid screening of candidate compounds.\cite{claussen2020detection,cao2020artificial,liu2021screening,schleder2021machine,andrejevic2022machine,ma2023topogivity,xu2024discovering} Models such as gradient-boosted trees trained on space group, electron count, and orbital-resolved valence descriptors have achieved strong performance,\cite{claussen2020detection} while neural networks applied to computed XANES spectra have further expanded predictive capabilities.\cite{andrejevic2022machine}. 

To address data and scalability constraints, composition-based heuristics—hereafter referred to as the physics-agnostic (PA) chemical rule—have been proposed, most notably the topogivity score $g^{PA}(M)$ introduced by Ma et al.,\cite{ma2023topogivity} which defines a learned elemental descriptor quantifying each element's intrinsic tendency to form topological phases. The topological character of a compound is then determined by the composition-weighted average of its constituent elements' topogivities, offering an interpretable, symmetry-independent criterion. Subsequent extensions have incorporated the Hubbard $U$ parameter for magnetic systems\cite{xu2024discovering} and quantum correlations across elements.\cite{xu2025quantum}

While PA chemical rules offer simplicity and interpretability, their predictive power is inherently constrained by their omission of critical physical factors that govern topological behavior. In real materials, nontrivial topology is shaped not only by elemental composition but also by crystal symmetry (space-group (SG)), electron filling, and relativistic effects such as spin–orbit coupling, particularly in systems containing heavy elements, as demonstrated in our recent work\cite{ullah2025txl}. By construction, PA chemical rules ignore these physically essential contributions. This omission not only limits their predictive accuracy but also leads to a fundamental ambiguity: such rules are intrinsically unable to distinguish between materials that share identical chemical compositions but crystallize in different space groups. Since topology is highly sensitive to symmetry and electronic structure, these compositionally identical yet structurally distinct materials can exhibit markedly different topological phases, rendering composition-only PA descriptors effectively agnostic to this distinction.

To overcome this limitation, we develop a physics-informed (PI) chemical rule $g^{PI}(M)$ that unifies compositional, chemical, and symmetry-derived information within an interpretable framework. Our approach integrates (i) elemental composition, enabling element-resolved topogivities; (ii) orbital and elemental category descriptors; and (iii) global constraints including electron filling and SG symmetry. These features are embedded in a linear model, yielding a topological score that explicitly decomposes into composition, chemical environment, and symmetry contributions.

This formulation retains the transparency of PA chemical rules while incorporating the physics necessary to resolve structure-dependent topology. High-throughput screening demonstrates that our approach achieves substantially enhanced accuracy compared to PA chemical rules, with balanced precision and recall for topological materials. Beyond accuracy gains, our framework resolves the degeneracy inherent to composition-only descriptors, enabling reliable discrimination between polymorphs with identical stoichiometry but distinct symmetry. By explicitly encoding symmetry and electronic structure, our PI chemical rule $g^{PI}(M)$ offers both enhanced predictive power and a physically transparent pathway for discovering topological materials in previously inaccessible regimes.

\section{Results}

\subsection{Physics-informed chemical rule}

Consider a material $ M$  composed of elements $ E$  with corresponding atomic fractions $f_E$ , satisfying $\sum_{E \in M} f_E = 1$ . Each material is represented by a feature vector $\mathbf{X}(M) \in \mathbb{R}^d$ , constructed as the concatenation of three physically motivated components: a composition block $ \mathbf{X}_c(M)$ , an auxiliary chemical feature block $\mathbf{X}_o(M)$ , and a global feature block $ \mathbf{X}_g(M)$ . The composition block encodes the fractional presence of elements excluding a reference element (oxygen in this work), thereby enabling a consistent parametrization of elemental contributions. The auxiliary chemical block captures local electronic and chemical environment effects, including orbital-resolved valence characteristics and element-category descriptors, while the global block incorporates physically essential constraints such as electron parity and SG symmetry through one-hot encoding.

A linear support vector classifier is trained on the full feature vector $\mathbf{X}(M)$  to classify materials into topological ($+1$)  or trivial ($-1$)  phases. The resulting decision function is given by
\begin{equation}
g^{PI}(M) = \mathbf{w} \cdot \mathbf{X}(M) + b,
\end{equation}
where $ \mathbf{w} = (\mathbf{w}_c, \mathbf{w}_o, \mathbf{w}_g)$  is the learned weight vector partitioned according to the three feature blocks, and $ b$  is a scalar bias. The predicted class is determined by the sign of $g^{PI}(M)$ , such that $ g^{PI}(M) > 0$  indicates a topological material, while $ g^{PI}(M) \leq 0$  corresponds to a trivial phase. Owing to the linear structure of the model, the decision function can be naturally decomposed into distinct physical contributions,
\begin{equation} \label{eq:gpi}
g^{PI}(M) = g_c(M) + g_o(M) + g_g(M),
\end{equation}
where each term corresponds to one of the feature blocks.

The compositional contribution takes the form
\begin{equation}
g_c(M) = \sum_{E \neq \mathrm{O}} f_E , \tau^c_E,
\end{equation}
where the quantity $ \tau^c_E = w^c_{E} + b$  defines the elemental score associated with element $E$ . This term recovers a composition-based PA chemical rule in which the topological tendency of a material is approximated as a weighted average of intrinsic elemental contributions. In this sense, the present formulation generalizes and embeds previous composition-only descriptors within a broader framework. The second contribution,
\begin{equation}
g_o(M) = \mathbf{w}_o \cdot \mathbf{X}_o(M),
\end{equation}
accounts for chemical environment effects beyond stoichiometry, including orbital hybridization and bonding characteristics that influence the electronic structure. The third contribution,
\begin{equation}
g_g(M) = \mathbf{w}_g \cdot \mathbf{X}_g(M),
\end{equation}
captures global physical constraints, most notably symmetry and electron parity, which play a decisive role in determining band topology.

Combining these terms, the final expression for the decision function becomes
\begin{equation}
g^{PI}(M) = \sum f_E \cdot  \tau^c_E + \mathbf{w}_o \cdot \mathbf{X}_o(M) + \mathbf{w}_g \cdot \mathbf{X}_g(M),
\end{equation}
which constitutes the proposed PI chemical rule. Unlike PA chemical rule (composition-only), this formulation is capable of distinguishing materials with identical chemical compositions but different SG symmetries or electronic configurations. At the same time, it preserves interpretability through the elemental topogivities $\tau^c_E$ , while systematically incorporating additional physically meaningful corrections. In this way, the proposed rule provides a unified and extensible framework that bridges chemical intuition and symmetry-based topological diagnostics, enabling more accurate and physically consistent identification of topological materials.

\subsection{Data-driven rationale for topological feature selection}

As reflected in Eq.~\eqref{eq:gpi}, our PI chemical rule \(g^{PI}(M)\) incorporates orbital characteristics, elemental categories, and global constraints such as electron filling and SG symmetry. The inclusion of these features is not arbitrary; rather, it is grounded in a systematic data-driven analysis that reveals clear quantitative trends linking these descriptors to topological behavior.

To establish this foundation, we analyze a large-scale dataset drawn from the topological materials database,\cite{topo_materials, bilbao_cryst, bradlyn2017topological, vergniory2019complete, vergniory2022all} comprising 38,184 materials computed using density functional theory with spin–orbit coupling. In this study, topological insulators and topological semimetals are treated as a single unified class of topological materials, resulting in 20,094 topological compounds (combining 6,109 TIs and 13,985 TSMs, ~52.6\%) and 18,090 trivial compounds (~47.3\%).

Starting from a broad pool of candidate descriptors—including chemical bonding characteristics, spin–orbit coupling strength (\(\propto Z^4\)), periodic table positions, electron counts, space group, valence electrons, and atomic mass—we perform an iterative feature selection process guided by both physical insight and statistical significance. This procedure yields a compact set of features that are both physically interpretable and predictive.

A key outcome of our analysis is the identification of electron filling as a primary factor governing topological character. As summarized in Table~\ref{tab:stat-analysis}, topological materials exhibit a substantially higher fraction of odd-electron systems (53.7\%) compared to trivial compounds (4.3\%), with topological semimetals showing the most pronounced enrichment (70.7\%). Conversely, trivial materials are overwhelmingly dominated by even-electron configurations (95.7\%), while this proportion drops to 46.3\% in the topological class. This pronounced imbalance reflects the fundamental role of band filling: odd electron counts naturally favor metallic or semimetallic states, which are closely associated with gapless or near-gapless topological phases.

Beyond electron filling, the distribution of valence orbitals further underscores the importance of orbital character. Topological materials show a strong enhancement in $d$- and $f$-orbital contributions (average occupations of 2.39 and 0.79, respectively) compared to trivial compounds (0.81 and 0.13), while $p$-orbital contributions are significantly reduced (1.25 vs. 2.47). This redistribution from $p$-dominated to $d$/$f$-dominated electronic character reflects the increasing importance of transition-metal and rare-earth elements, whose strong spin–orbit coupling and enhanced hybridization are central to driving band inversion.

The distribution of elemental categories provides complementary chemical insight. Topological materials are strongly enriched in transition metals (35.6\% vs. 11.9\%) and lanthanides (10.1\% vs. 2.1\%), while trivial compounds are dominated by nonmetals (47.4\% vs. 19.6\%) and halogens (11.9\% vs. 3.0\%). These systematic differences reflect a shift from ionic, localized bonding in trivial systems toward delocalized, metallic bonding in topological materials—environments that naturally favor band inversion and nontrivial topology.

\begin{figure}
    \centering
    \includegraphics[width=1.0\linewidth]{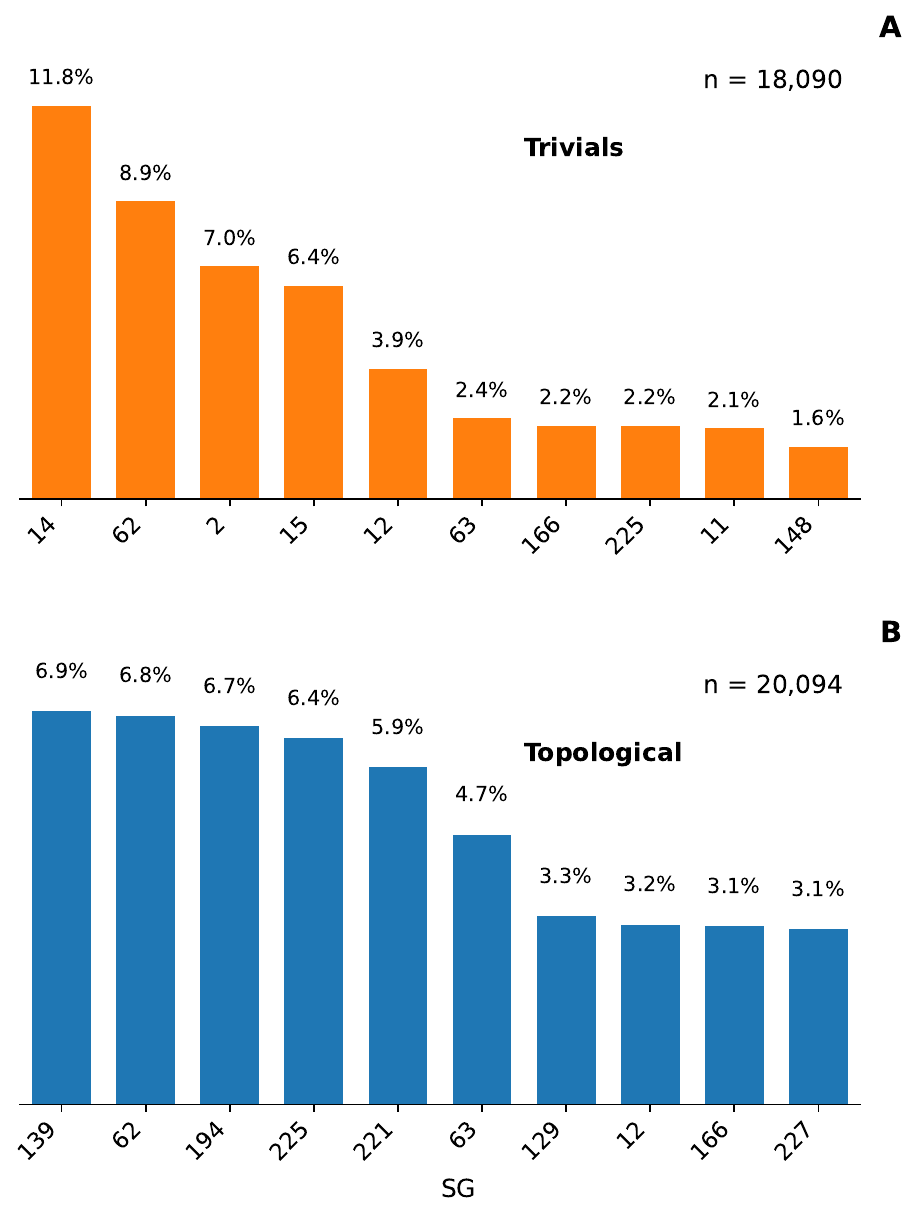}
    \caption{SG distribution of (A) trivial, (B) topological materials where $n$ represents the total number of compounds of a class.}
    \label{fig:sg_dist}
\end{figure}

Finally, SG symmetry emerges as a key determinant, exhibiting clear and systematic differences between topological and trivial materials. As shown in Fig.~\ref{fig:sg_dist}, among the 216 SGs represented, trivial compounds are predominantly associated with lower-symmetry (monoclinic and orthorhombic) groups, including 14 (11.8\%), 62 (8.9\%), 2 (7.0\%), and 15 (6.4\%). In contrast, topological materials display pronounced peaks in higher-symmetry groups such as 139 (6.9\%), 62 (6.8\%), 194 (6.7\%), 225 (6.4\%), and 221 (5.9\%), which are well known to support symmetry-enforced topological phases. 

This symmetry-driven distinction is further reflected in the exclusive presence of certain space groups within each dataset. Groups such as 196, 103, 106, 175, 210, and 211 appear only among topological materials, while over 30 space groups—including 3, 16, 17, 145, 22, 151, 24, 153, 27, 32, 37, 39, 169, 42, 45, 48, 49, 177, 183, 192, 195, 77, 78, 80, 81, 208, 94, 98, 105, and 112—are unique to the trivial set. These compositional asymmetries confirm that SG information is a critical factor in distinguishing topological from trivial materials.

Together, these data-driven observations demonstrate that topological behavior is governed by a combination of electron filling, orbital character, chemical environment, and crystallographic symmetry—factors not captured by composition alone. This motivates our PI chemical rule, which explicitly integrates these features into a unified and interpretable framework.

\begin{table}[htbp] 
\centering
\caption{Comparative statistical analysis between trivial and topological materials.}
\label{tab:stat-analysis}
\begin{threeparttable}
\begin{tabular}{@{}>{\raggedright}p{3.5cm}lS[table-format=1.4]S[table-format=1.4]@{}}
\toprule
\textbf{Category} & \textbf{Subcategory} & \textbf{Trivial} & \textbf{Topological} \\
\midrule

\multicolumn{4}{@{}l}{\textbf{Distribution (\%)}} \\
\cmidrule(l){2-4}
& Odd & 4.3 & 53.7 \\
& Even & 95.7 & 46.3 \\
\addlinespace

\multicolumn{4}{@{}l}{\textbf{Element Ratios (\%)}} \\
\cmidrule(l){2-4}
& Nonmetal & 47.4 & 19.6 \\
& Halogen & 11.9 & 3.0 \\
& Transition metal & 11.9 & 35.6 \\
& Alkali metal & 7.9 & 2.8 \\
& Metalloid & 8.3 & 12.8 \\
& Metal & 5.4 & 10.2 \\
& Alkaline earth metal & 4.5 & 4.0 \\
& Lanthanide & 2.1 & 10.1 \\
& Actinide & 0.6 & 1.8 \\
& Noble gas & 0.1 & 0.1 \\
\addlinespace

\multicolumn{4}{@{}l}{\textbf{Average valence electrons}} \\
\cmidrule(l){2-4}
& s & 1.82 & 1.81 \\
& p & 2.47 & 1.25 \\
& d & 0.81 & 2.39 \\
& f & 0.13 & 0.79 \\
\addlinespace
\bottomrule
\end{tabular}

\begin{tablenotes}
\small
\item Note: Percentages may not sum to 100\% due to rounding.
\end{tablenotes}
\end{threeparttable}
\end{table}

\subsection{Performance evaluation}

In this section, we assess the capability of the proposed PI chemical rule $g^{PI}(M)$ to distinguish topological materials from trivial ones and benchmark it against the composition-based PA chemical rule $g^{PA}(M)$. The dataset (sources from topological materials database) was divided into 80\% for training and 20\% for testing (7637 materials), with the latter designated as Discovery Space-1, on which high-throughput screening was conducted using features learned from the training data. Model performance is evaluated using precision, recall, and F1-score, as defined in Eq.~\eqref{eq:metrics}:

\begin{equation}\label{eq:metrics}
\text{Precision} = \frac{TP}{TP + FP}, \quad
\text{Recall} = \frac{TP}{TP + FN}, \quad
\text{F1} = 2 \cdot \frac{\text{Precision} \cdot \text{Recall}}{\text{Precision} + \text{Recall}} ,
\end{equation}

where $TP$, $FP$, and $FN$ denote true positives, false positives, and false negatives. Precision reflects the fraction of predicted positives that are correct, recall measures the fraction of true positives recovered, and F1-score balances the two.

As summarized in Table~\ref{tab:discovery1_comparison}, our PI chemical rule $g^{PI}(M)$ achieves a significantly higher overall accuracy of 0.87, compared to 0.82 for the PA chemical rule $g^{PA}(M)$ —a clear demonstration of the value added by incorporating physics-based descriptors. The improvement is consistent across both classes. For topological materials, our rule achieves precision of 0.88 and recall of 0.87, yielding an F1-score of 0.87. In contrast, the PA chemical rule $g^{PA}(M)$  manages only 0.87 precision and 0.77 recall for the same class, with a lower F1-score of 0.82. This indicates that our $g^{PI}(M)$ chemical rule not only identifies topological candidates more reliably but also captures a substantially larger fraction of true topological materials. For trivial materials, our rule similarly outperforms the original, delivering higher precision (0.85 vs. 0.77) while maintaining comparable recall (0.87).

In addition, a notable distinction lies in the performance balance between classes. The $g^{PA}(M)$ chemical rule exhibits asymmetric behavior, achieving high recall for trivial materials (0.87) but substantially lower recall for topological ones (0.77), suggesting it is biased toward identifying trivial phases. In contrast, our chemical rule achieves balanced performance across both classes, with recall values of 0.87 for both trivial and topological materials. This symmetry reflects the strength of our physics-guided design, which encodes chemically and crystallographically meaningful descriptors that are equally informative for both classes.

\begin{figure}
    \centering
    \includegraphics[width=1.0\linewidth]{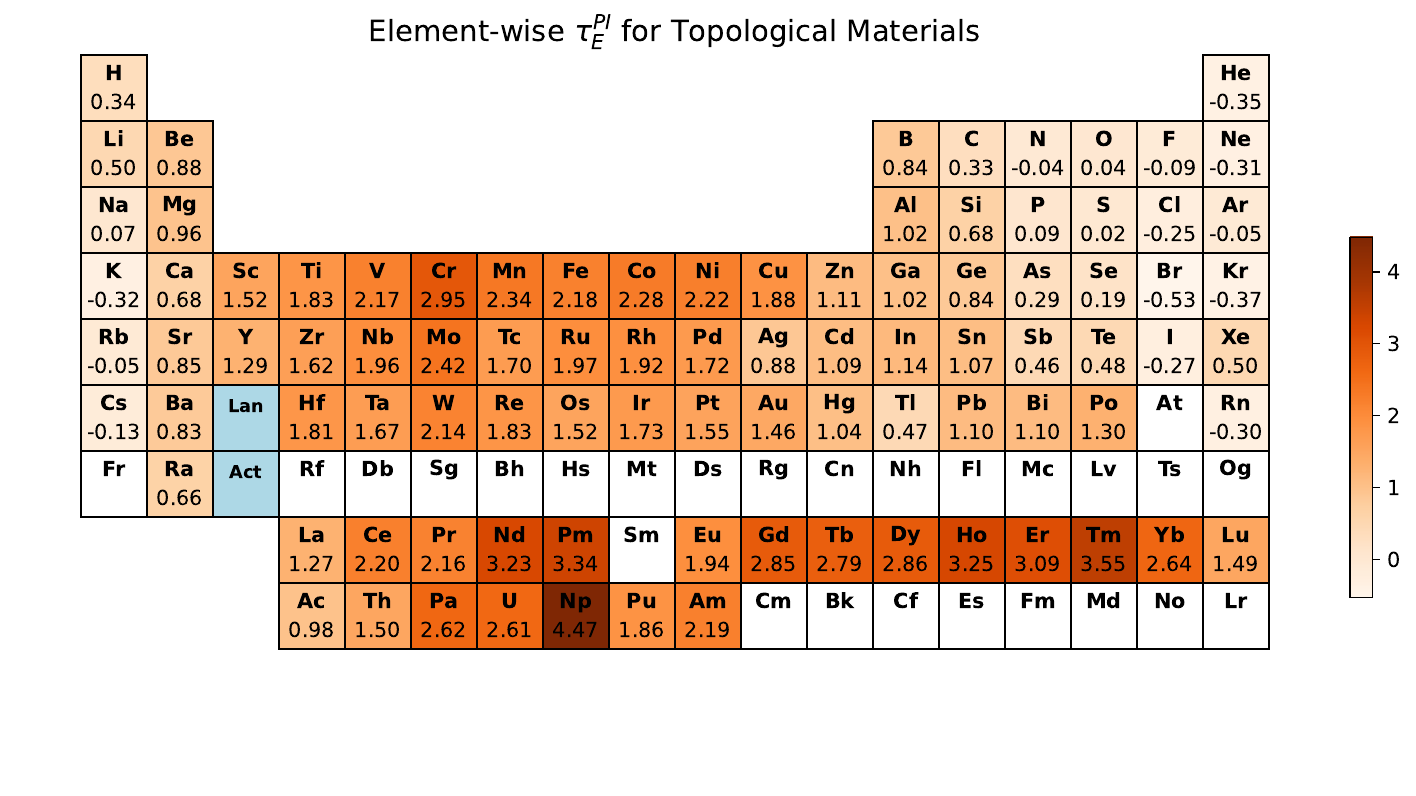}
    \caption{Periodic table distribution of PI topogivities $\tau^{PI}_E$ for different elements. Elements are colored according to their $\tau^{PI}_E$ values, with warmer colors indicating higher propensity to form topological phases. Strong spin–orbit coupling elements (e.g., Fe, Co, Ni, Cu) show elevated scores, while halogens, alkali metals, and noble gases exhibit negative values associated with trivial behavior.}
    \label{fig:tau_pt}
\end{figure}

Furthermore, to provide chemical interpretability, we present the periodic table distribution of our PI topogivities $\tau^{PI}_E$ in Figure~\ref{fig:tau_pt}, which integrate valence electron configuration and elemental category into the learned elemental descriptor. These values are derived from the linear decomposition:

\begin{equation}
\tau^{PI}_E = \tau^c_E + w_{\text{orb}} \cdot o_E + w_{\text{cat}} \cdot k_E,
\end{equation}

where $\tau^c_E$ is the composition-only topogivity, $o_E$ represents normalized valence orbital occupations, $k_E$ is a one-hot vector encoding the elemental category, and $w_{\text{orb}}$, $w_{\text{cat}}$ are the corresponding learned weights.

The resulting $\tau^{PI}_E$ values align well with established chemical intuition. Elements with strong spin–orbit coupling—such as Fe, Co, Ni, and Cu—receive high scores, consistent with their prevalence in topological materials. Conversely, lighter elements like H, C, P, and S exhibit low values, while halogens, several alkali metals, and most noble gases yield negative scores, reflecting their strong association with trivial band structures.

\begin{table}[htbp]
\centering
\caption{Performance comparison between the PI chemical rule $g^{PI}(M)$ and the composition-based PA chemical rule $g^{PA}(M)$ on Discovery Space–1.}
\label{tab:discovery1_comparison}
\begin{tabular}{lcccc}
\toprule
\textbf{Model} & \textbf{Class} & \textbf{Precision} & \textbf{Recall} & \textbf{F1-score} \\
\midrule
\multirow{2}{*}{$g^{PA}(M)$} 
& Trivial       & 0.77 & 0.87 & 0.82 \\
& Topological   & 0.87 & 0.77 & 0.82 \\
\cmidrule(lr){1-5}
& \textbf{Accuracy} & \multicolumn{3}{c}{0.82} \\
\midrule
\multirow{2}{*}{$g^{PI}(M)$} 
& Trivial       & 0.85 & 0.87 & 0.86 \\
& Topological   & 0.88 & 0.87 & 0.87 \\
\cmidrule(lr){1-5}
& \textbf{Accuracy} & \multicolumn{3}{c}{0.87} \\
\bottomrule
\end{tabular}
\end{table}

\begin{figure}
    \centering
    \includegraphics[width=1.0\linewidth]{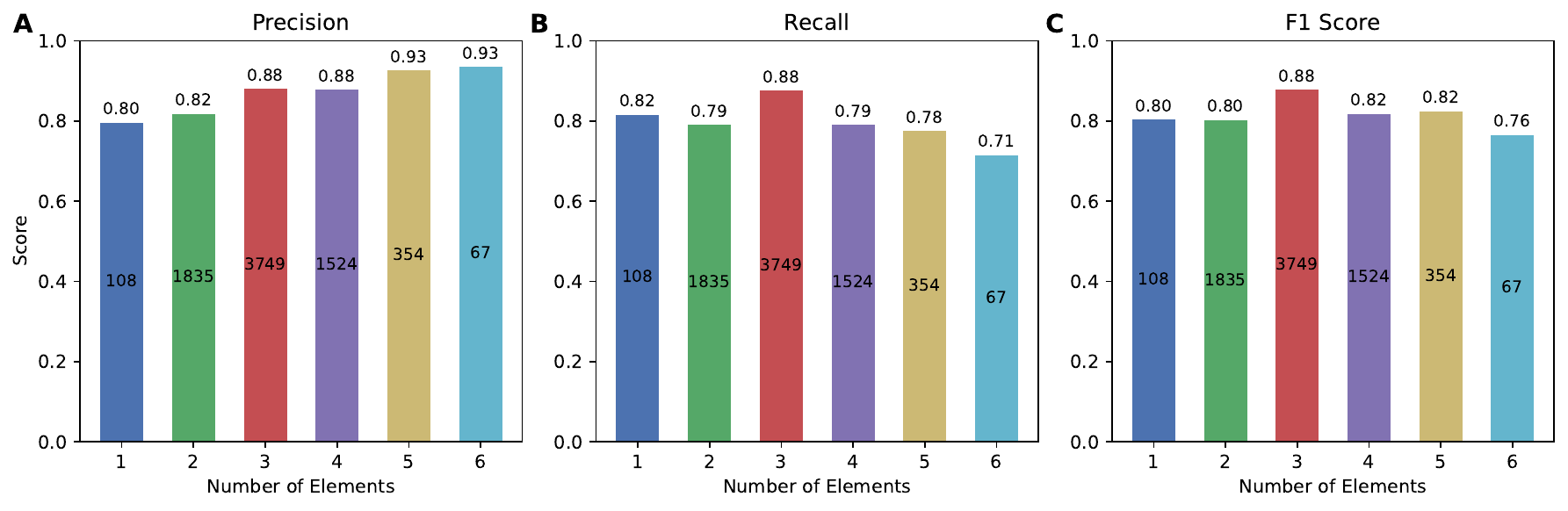}
    \caption{Performance metrics of the $\tau^{PI}$ chemical rule on Discovery Space‑1 stratified by the number of constituent elements. (A) Precision, (B) Recall, and (C) F1‑score for materials containing 1 to 6 elements. The number on the top of each bar indicates the average metric value, and the number inside each bar denotes the number of test samples for that category.}
    \label{fig:ele_per_pt}
\end{figure}

\textbf{Performance across compositional complexity:} To systematically assess the model's generalization across varying chemical complexity, we evaluate its performance on Discovery Space–1 (7,637 samples) stratified by the number of constituent elements. The distribution spans from unary to senary compounds, with counts of [108, 1,835, 3,749, 1,524, 354, 67], respectively. Figure~\ref{fig:ele_per_pt} summarizes the corresponding precision, recall, and F1-scores for each subset.

Ternary systems dominate the dataset, comprising 3,749 samples (~49\%) in Discovery Space–1 and 15,180 samples (~48\%) in the training set. Unsurprisingly, the model achieves optimal performance on this class, with precision, recall, and F1-score all reaching 0.88. This near-perfect balance reflects the statistical prevalence of ternary compounds and, more importantly, suggests that the underlying structure–property relationships governing topological phases are well-represented and learnable within this chemical subspace.

For higher-order systems—quaternary through senary—the situation changes dramatically due to severe class imbalance, driven by the progressive scarcity of topological samples. In the training set, the fraction of topological materials drops sharply to 27.2\%, 25.2\%, and 11.5\% for four-, five-, and six-element compounds, respectively, with analogous trends observed in Discovery Space–1. Despite this imbalance, the model exhibits remarkably robust precision: 0.88, 0.93, and 0.93 for quaternary, quinary, and senary classes. Such high specificity indicates that the learned decision boundary is highly conservative, effectively rejecting false positives even in data-sparse regimes.

However, this comes at the cost of recall, which falls to 0.79, 0.78, and 0.71, respectively. This precision-recall trade-off is characteristic of classifiers trained on imbalanced data: the model prioritizes reliability over coverage when positive examples are scarce, inevitably missing a nontrivial fraction of true topological materials in higher-order chemical spaces.

Critically, the training-to-test ratios for these subsets ($\approx$3.91, 4.14, and 3.49) are comparable to that of ternary compounds ($\approx$4.05), ruling out test set sampling bias as the explanation for degraded performance. Instead, the limiting factor is the absolute scarcity of topological instances in higher-order systems. The diminished F1-score for complex compositions therefore reflects a data bottleneck rather than a fundamental limitation of the physics-informed chemical rule in capturing higher-order chemical interactions.

\textbf{Distinguishing polymorphs via SG encoding:}
Beyond the overall and element‑wise performance, a key advantage of the $g^{PI}(M)$ chemical rule is its ability to distinguish materials with identical compositions but different crystal structures. The PA (physics-agnostic) chemical rule $g^{PA}(M)$, relying purely on composition, assigns identical $g$ values to such systems and therefore cannot resolve structural effects. In contrast, our approach incorporates crystallographic symmetry as a core descriptor, enabling structure-sensitive predictions. For example, Re$_1$N$_2$ in space groups 71 and 62 receives the same value from the $g^{PA}(M)$ $(-0.1355)$, whereas the PI chemical rule $g^{PI}(M)$ clearly differentiates them $(1.8488$ and $2.5111)$. Similarly, Mo$_9$Se$_{11}$ (SG 63 vs. 176) is indistinguishable under the $g^{PA}(M)$ $(0.6983)$ but separated by our model $(1.5430) \text{ vs. } (2.9589)$. In the case of La$_1$Ni$_1$O$_3$, the $g^{PA}(M)$ fails to capture the difference between polymorphs (SG 99 vs. 221), while the our chemical rule correctly reflects their contrasting behavior ($0.1878$ vs. $-0.2718$).

\textbf{Assessing generalizability on unseen materials:}
To further evaluate the predictive capability of the our $g^{PI}(M)$ chemical rule, we explored a challenging discovery space consisting of 1,433 materials reported by Ma et al.\cite{ma2023topogivity}, originally curated by Tang et al.\cite{tang2019comprehensive}. This dataset is particularly significant because the topological character of these materials cannot be determined using conventional symmetry indicators, making them an ideal benchmark for testing models that incorporate chemical and electronic descriptors beyond symmetry. Among these compounds, 1,235 are already included in the topological materials database, leaving 198 previously uncharacterized candidates. We designate this subset as Discovery Space-2, representing a stringent test set for out-of-distribution generalization.

Applying our chemical rule to these 198 candidates, we identified 12 materials as potential topological phases with varying confidence levels. Notably, several compounds exhibit strong positive scores ($g^{PI}(M) > 1$), indicating high-confidence predictions. These include Ag$_1$Pb$_4$Pd$_6$ (SG 152, $g^{PI}(M)=2.34$), Ta$_{21}$Te$_{13}$ (SG 183, $g^{PI}(M)=1.74$), O$_1$Ti$_6$ (SG 159, $g^{PI}(M)=1.29$), In$_1$Sr$_1$ (SG 43, $g^{PI}(M)=1.17$), and La$_1$Mo$_2$O$_5$ (SG 186, $g^{PI}(M)=1.02$). The consistently large positive $g^{PI}(M)$ values for these materials suggest robust topological character as inferred from the learned chemical rule. In addition, several compounds such as P$_3$Sc$_7$ (SG 186, $g^{PI}(M)=0.72$) and Al$_{22}$Mo$_5$ (SG 43, $g^{PI}(M)=0.42$) show moderate confidence, indicating proximity to the topological decision boundary. A smaller subset, including Ag$_2$P$_1$S$_3$ (SG 19) and Bi$_1$Se$_3$Sr$_1$ (SG 19), yields marginally positive scores ($g^{PI}(M) \approx 0$), reflecting weaker or borderline topological tendencies. These cases may be sensitive to subtle structural or electronic details not fully captured by the current descriptors.

\section{Discussion}

We have developed a PI (physics-informed) chemical rule that combines compositional features with orbital character and space-group symmetry within an interpretable framework. By explicitly incorporating electron filling and symmetry constraints, the model overcomes a key limitation of PA (physics-agnostic) composition-only approaches, enabling discrimination between polymorphs and capturing structure-dependent topological behavior.

Our results demonstrate superior and balanced performance against the PA chemical rule proposed by Ma et al.\cite{ma2023topogivity} on a discovery space of 7,637 materials, underscoring the model's potential as a reliable and generalizable tool for topological materials discovery. Analysis across compositional complexity further reveals that the model maintains strong predictive specificity even in chemically complex regimes. A notable strength of the approach is its ability to resolve structural degeneracy through explicit symmetry encoding, capturing symmetry-driven electronic effects that are inaccessible to composition-only descriptors.

To evaluate generalizability on unseen materials, we tested our PI chemical rule on a completely independent dataset of 198 previously uncharacterized compounds—materials whose topological character cannot be resolved using conventional symmetry indicators. From these challenging candidates, the model identified 12 potential topological phases with varying confidence levels, demonstrating its utility in regimes where symmetry-based approaches fail.

In conclusion, the physics-informed chemical rule provides a transparent, data-efficient, and physically interpretable approach to topological materials discovery. By bridging heuristic rules and black-box machine learning, it offers a unified framework that leverages both domain knowledge and statistical learning. Future work will focus on alleviating data sparsity in complex compositional spaces, extending the framework to magnetic and symmetry-broken systems, and integrating with first-principles calculations toward closed-loop, autonomous discovery of quantum materials.

\section{Methods}
\subsection{Data and Materials}
We sourced our data from the topological materials database,\cite{topo_materials, bilbao_cryst, bradlyn2017topological, vergniory2019complete, vergniory2022all} which contains density functional theory (DFT) calculations with spin–orbit coupling for 38,184 materials. For model training and validation, we used a curated subset of 7,637 materials (Discovery Space–1), with a separate test set of 198 previously uncharacterized compounds (Discovery Space–2) drawn from the compilation by Tang et al.\cite{tang2019comprehensive} and later used by Ma et al.\cite{ma2023topogivity} Topological insulators and topological semimetals were treated as a single unified class of topological materials.

\subsection{Feature engineering}
Our PI chemical rule integrates three distinct feature blocks: compositional, chemical, and global symmetry-based descriptors.

\textbf{Compositional block.} Elemental composition is encoded using fractional atomic concentrations. To enable a consistent parameterization, oxygen is selected as the reference element and excluded from the composition vector. The remaining elements are represented by their fractional abundances relative to the total atom count.

\textbf{Chemical descriptors.} For each compound, we compute orbital-resolved valence features. Using the electronic structure of each element, we extract the occupation numbers of s, p, d, and f valence orbitals, normalized by their respective maximum occupancies (2, 6, 10, and 14). These are averaged over all elements weighted by their fractional concentrations. Additionally, each element is assigned to one of eleven coarse-grained categories (e.g., transition metal, lanthanide, nonmetal), and a category frequency vector is constructed.

\textbf{Global constraints.} Two global descriptors are included: electron filling parity (odd vs. even total electron count) and crystallographic symmetry encoded as a one-hot vector over space groups. This enables the model to capture symmetry-protected band degeneracies and to distinguish between compositionally identical polymorphs.

\subsection{Model architecture and training}
We employed a linear support vector classifier (LinearSVC) as the base model due to its interpretability and efficiency with high-dimensional sparse features. The three feature blocks—compositional, chemical, and global--are concatenated into a single feature vector. The model is trained to classify materials as topological (label +1) or trivial (label -1) using a hinge loss objective with L2 regularization. Training was performed on the training set (30,547 materials) with a maximum of 10,000 iterations to ensure convergence.

\subsection{Extraction of PI topogivities}
A key output of our framework is a set of element-specific PI topogivities $\tau^{PI}_E$, which quantify each element's intrinsic tendency to host topological phases. These are derived from the trained linear model as follows.

The composition block is parameterized with oxygen as the reference element. Consequently, the model intercept $b$ corresponds to the contribution of oxygen. For any other element $E$, its compositional elemental score is given by $\tau^{c}_E = w_E + b$, where $w_E$ is the learned weight for that element's fractional concentration.

To obtain the full PI topogivity, we further project the chemical descriptor weights onto each element. For an element $E$, we construct its element-specific chemical feature vector (normalized orbital occupations and category one-hot encoding) and compute the dot product with the learned weights $w_{\text{other}}$. This contribution is added to the compositional topogivity, yielding the complete physics-informed value:

\begin{equation}
\tau^{PI}_E = w_E + b  + w_{\rm other} \cdot \mathbf{x}_E,
\end{equation}
where $\mathbf{x}_E$ is the chemical feature vector for element $E$.

\subsection{Model evaluation and inference}

For a given test compound, the topological score $g^{PI}(M)$ is computed as:

\[
g^{PI}(M) = \sum f_E \cdot \tau^c_E   + \text{(chemical terms)} + \text{(global terms)},
\]

where $f_E$ is the fractional concentration of element $E$. Materials with $g^{PI}(M) > 0$ are classified as topological; those with $g^{PI}(M) \leq 0$ are classified as trivial. Performance was evaluated using precision, recall, F1-score, and overall accuracy, stratified by the number of constituent elements to assess generalization across chemical complexity.
\begin{acknowledgement}

A.U. acknowledges funding from the National Natural Science Foundation of China (No. W2433037) and the Natural Science Foundation of Anhui Province (No. 2408085QA002). M. Y. acknowledges funding from Scientific Research Projects of Anhui Provincial Department of Education under Grant Nos. 2025AHGXZK20126 and 2025AHGXZK50054.

\end{acknowledgement}

\section{Code and Data Availability}

The code used for feature engineering, model training, and evaluation will be made available upon acceptance of this paper at \href{https://github.com/Arif-PhyChem/physics_informed_chemical_rule}{https://github.com/Arif-PhyChem/physics\_informed\_chemical\_rule}.


\bibliography{ref}

\end{document}